# Multi-spectral programmable absorbers

Yun Meng [1], Dan Li [1], Chong Zhang [3], Yang Wang [3], Robert E. Simpson [2*], Yi Long [1*]

[1] School of Materials Science and Engineering, Nanyang Technological University, 50 Nanyang Avenue, 639798, Singapore

[2] ACTA Lab, Singapore University of Technology and Design, 8 Somapah Road, 487372, Singapore

[3] Shanghai Institute of Optics and Fine Mechanics, Chinese Academy of Sciences, Shanghai 201800, China

Corresponding author: robert_simpson@sutd.edu.sg and longyi@ntu.edu.sg

## Abstract

We designed and demonstrated a multi-spectral programmable perfect absorber that exploits two different phase-change materials. This programmability is possible by resonantly coupling two phase change materials, a $Ge_2Sb_2Te_5$ layer to vanadium dioxide nanoparticles ($VO_2$ NPs). The perfect absorption is attributed to the coalescence of gap plasmon modes excited between the NPs and waveguide cavity-like modes excited between the film and the NPs. The absorptance peak (>90%) can be tuned to four different infrared (IR) wavelengths from 1906 to 2960 nm by heating the structure to different temperatures. The perfect absorber is reconfigurable, lithography-free, large-scale, polarization-insensitive omnidirectional. Our strategy opens a new path for programmable infrared photonics.

Corresponding author: robert_simpson@sutd.edu.sg; longyi@ntu.edu.sg

Perfect electromagnetic (EM) absorption has wide ranging applications, such as biosensing, display, photo-thermal conversion. [1-6] However, it is hard to achieve perfect absorption using natural materials due to the impedance-mismatch problem. Therefore, several plasmonic nanostructures and metasurfaces have been proposed to achieve high absorptance. [7-11] However, the optical response of these devices is usually fixed once the designed metasurface has been fabricated.

To enable tunable and reconfigurable perfect absorbers, there are essentially two options: (1) tune the geometry of the resonant cavity, or (2) tune the refractive index of the resonant cavity. Micro-electromechanical systems can be used to tune the cavity geometry, but this require exquisite control of the micropatterning conditions and the devices can have cyclability issues. [12,13] On the other hand, materials that exhibit a tunable refractive index, such as phase-change materials, graphene [14,15] and liquid crystals [16] are a viable solution. In particular, the phase change material. $Ge_2Sb_2Te_5$ (GST) incorporated metamaterials have shown promise for tuning of plasmon resonances. This tunability is attributed to plasmon resonances being sensitivity to the refractive index ($n$) of the surroundings. Since the $n$ of $Ge_2Sb_2Te_5$ can reversibly change from ~4 to ~6 in the IR between the amorphous state and crystalline states, using laser or electrical heat pulses [17-19], it is particularly attractive for M-IR reprogrammable plasmonics. [20-25]

Patterning the nanoscale features of a metasurface usually requires expensive and time-consuming techniques, such as electron beam lithography. These points make electron beam lithography unattractive mass fabrication on an industrial scale. In contrast, metallic nanoparticles (NPs) that exhibit local surface plasmon resonances (LSPRs) can be manufactured at a low-cost synthesized.[26,27] We propose, therefore, to couple these plasmonic phase change NPs with other phase change plasmonic structures to achieve electromagnetic resonances that can be tuned across a broadband.[28,29] By combining multiple phase change materials, multiple resonant modes can be achieved.

In our previous work, we combined $Ge_2Sb_2Te_5$ and vanadium dioxide ($VO_2$) thin films to achieve a 2-bit, 4-state reflective switch. The switch allowed four different reflectivity levels for light with a wavelength of 680 nm.[30] We now extend this work by combining $VO_2$ nanoparticles (NPs) with a $Ge_2Sb_2Te_5$ thin film structure. $VO_2$ NPs undergo a metal-insulator-transition that changes the LSPRs wavelength.[31-34] We hypothesize that combining $VO_2$ NPs and $Ge_2Sb_2Te_5$ films can provide for different resonant wavelengths. We demonstrate a 4-state tunable resonant wavelength shift of a perfect absorber using self-assembled $VO_2$ NPs coupled to $Ge_2Sb_2Te_5$ thin films. The reversible change of the 4-state absorptance peaks allows perfect absorption programming from 1733 nm to 2888 nm. The LSPRs between the metals create a strong magnetic dipolar moment and thus achieve a high absorption. Furthermore, the strong absorptance is independent of the incident angle and the light's polarization state. The polarization-insensitive, large-scale, lithography-free and multi-level tunable functionality provides a fine modulation on the plasmonic resonant modes This work lays the foundation for smart sensing, programmable modulators, reconfigurable optical switches, and more.

A schematic of the four-wavelength switch is shown in Figure 1(a). It consists of an 80 nm thick $Ge_2Sb_2Te_5$ layer, which was deposited on a silicon substrate by radio frequency sputtering (AJA Orion5) from a $Ge_2Sb_2Te_5$ alloy target. A TiN layer, which was deposited using a Ti target by direct current magnetron sputtering in a reactive nitrogen atmosphere. The $N_2$ and Ar gases had the ratio 1:4. The Au film with a thickness of 90 nm was sputtered using a Au target with an RF power of 100 W. The chamber's background and sputtering pressure were $4 \times 10^{-5}$ Pa and 0.2 Pa, respectively. The average particle size of $VO_2$ NPs (Ji-cheng, China) was 45 nm. The $VO_2$ NPs were dispersed in alcohol with the centration of 2%. Then the $VO_2$ NPs solution was spin-coated onto the surface of the Au film.

The Finite-difference time-domain (FDTD, Lumerical) technique was used to solve Maxwell's equations and simulate the reflectance spectrum for different device design. In the model, the optical properties of $VO_2$ NPs and the $Ge_2Sb_2Te_5$ layer in Figs. 1e and 1f were used. The permittivity of TiN and Au were given by Philipp's data and Johnson and Christy's data. [35] The angular reflection of the absorber was measured using a spectroscopic ellipsometer for different incident angles varying from 20° to 80°. The absorptance was then calculated from A=1-R-T, where R and T are respectively the transmissivity and reflectivity. The size of $VO_2$ NPs was measured by transmission electron microscopy (TEM) in the bright field mode.

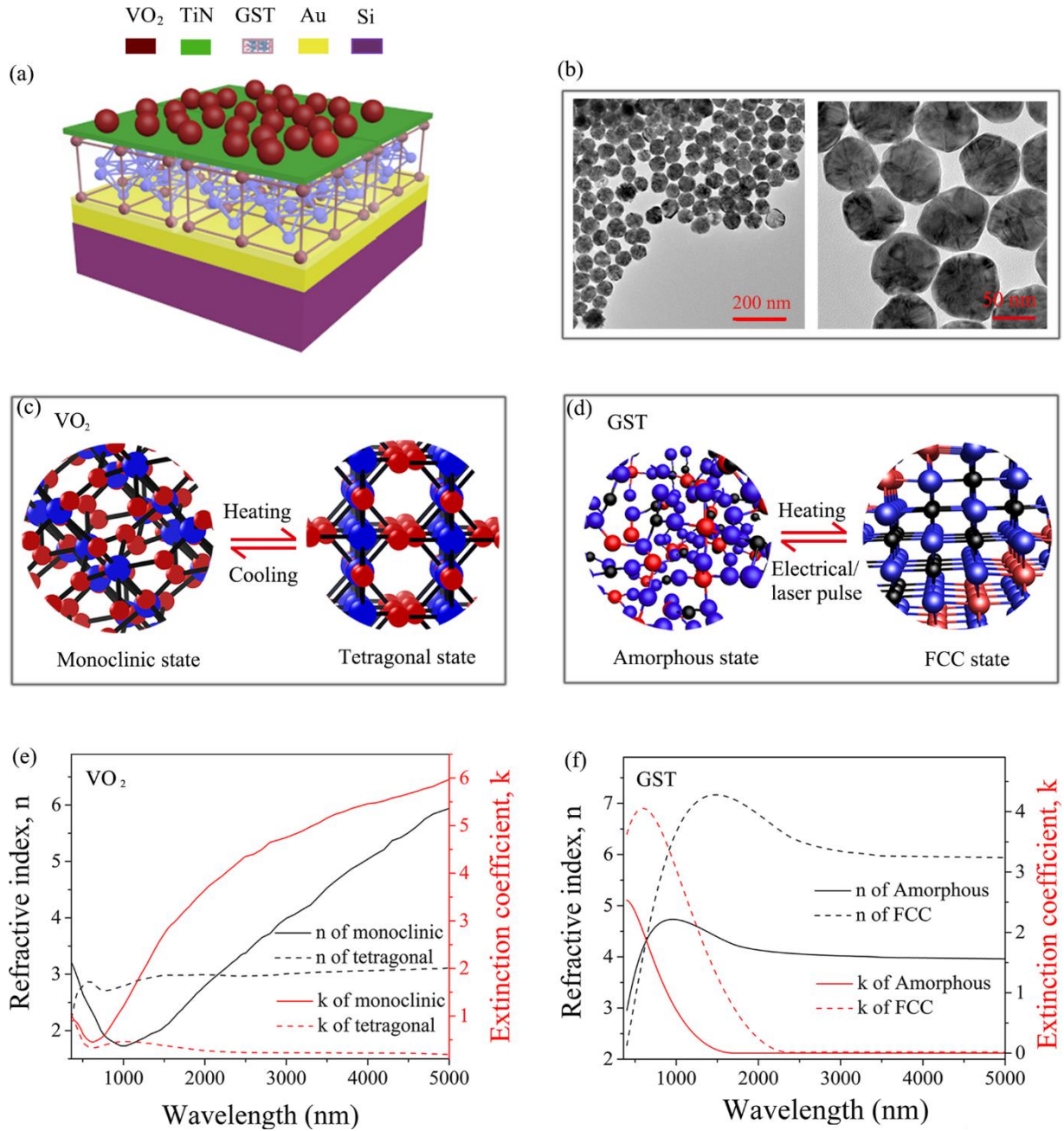

Figure 1 (a) Schematic of the reconfigurable absorber. The average diameter of VO$_2$ NPs is 90 nm. The thickness of TiN, GST and Au layer are 5 nm, 90 nm, and 80 nm, respectively. (b) TEM images of surface of the proposed absorber; (c) Schematic illustration of the reversible switching of VO$_2$ between monoclinic and tetragonal states; (d) Schematic illustration of the reversible switching of the GST between amorphous and Face Centered Cubic (FCC) states; The refractive index of (e) VO$_2$; (f) GST.

Fig. 1a illustrates the proposed reconfigurable absorber. An Au layer of 80 nm was firstly deposited on silicon (Si) substrate. The Au layer acts as a mirror to enhance the reflected signal, while also interacting with the top VO$_2$ NPs (tetragonal state) to create displacement current loops, which leads to the strong magnetic dipolar resonance. A 5 nm TiN was sandwiched between the $Ge_2Sb_2Te_5$ and $VO_2$ NPs to protect the $Ge_2Sb_2Te_5$ layer from oxidation [36,37] and interlayer diffusion [38]. The diameter of the $VO_2$ NPs and the thickness of GST were optimized by the aforementioned finite-difference time-domain simulations. Finally, the GST layer of 90 nm was sputtered and the $VO_2$ NPs with the radius of 45 nm were spin coated from a colloidal solution. The $VO_2$ NPs size was confirmed by transmission electron microscopy. Ab image of the particles is presented in Fig. 1b.

The reversible phase transitions of the $VO_2$ NPs and GST are accompanied by a reversible change to the optical properties. As shown in Fig. 1c, the monoclinic state of the $VO_2$ NPs displays a distorted structure with zigzag-type V–V atomic chains at room temperature and it changes to tetragonal state with straight V–V chains above ~65 °C. This change in structure causes the optical properties to change as evident by the refractive index shown in Fig. 1e. Similarly, Fig. 1d shows how the amorphous structure of $Ge_2Sb_2Te_5$ crystallizes into to face centered cubic (FCC) structure above 180 °C. Again, this phase transition is accompanied by an abrupt change to the refractive index, see Fig. 1f. Note, the re-amorphous state of $Ge_2Sb_2Te_5$ is achieved by quenching the material from a temperature above the melting temperature ($T_M$ = 873 K). This is possible by laser or electrical pulses.

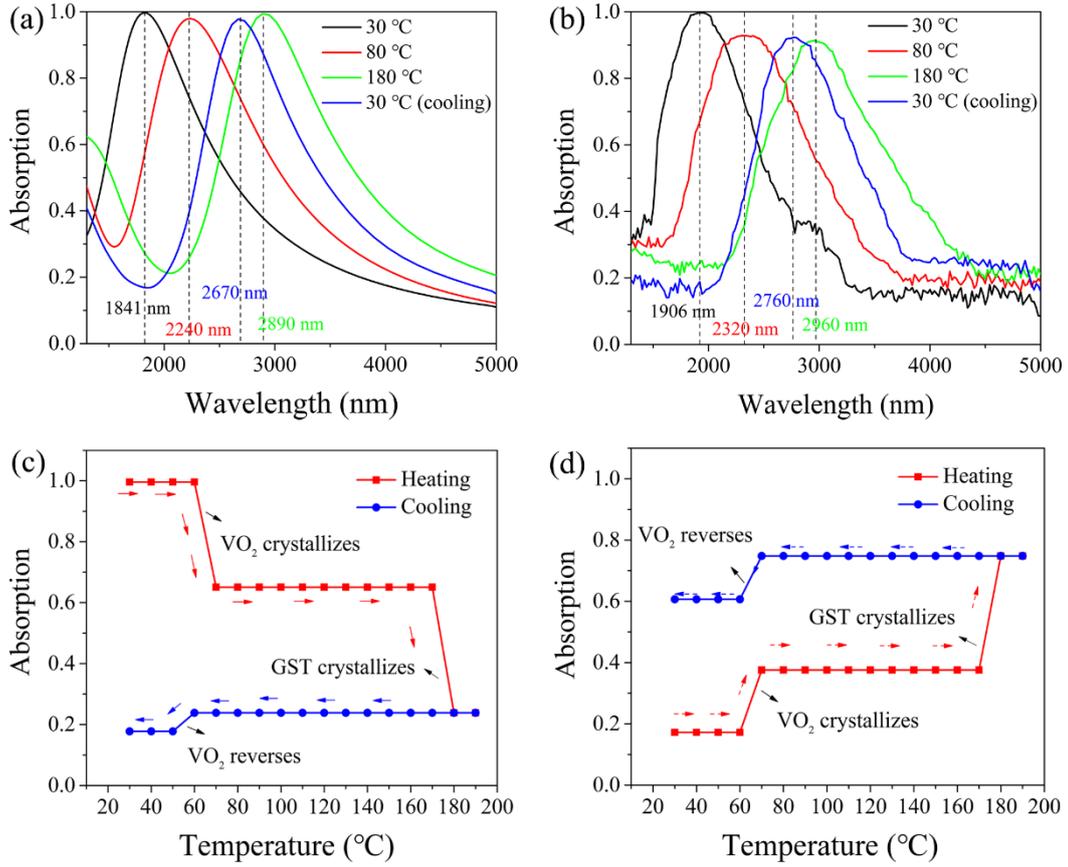

Figure 2 (a) Simulated and (b) measured absorption changes with wavelength o under 30 °C, 80 °C, 180 °C, and cooling back to 30 °C. The measured absorption changes with temperature at the wavelength of (c) 1900 nm and (d) 3300 nm.

Fig. 2 shows the simulated and measured absorption of the proposed structure. The absorption (A(λ)) is derived from A(λ) = 1 − T(λ) − R(λ). Since the transmittance spectrum (T(λ)) for the proposed absorber is zero, A(ω) = 1 − R(λ), where R(λ) is the reflectance spectrum of the absorber. The simulated absorption of the metasurface is presented in Fig. 2a. At room temperature, an absorption peak with A ($\lambda_1$=1841 nm) = 0.99 is achieved with the amorphous $Ge_2Sb_2Te_5$ layer and monoclinic $VO_2$ NPs. When temperature is raised to 80 °C, the resonant peak spectrally redshifts to $\lambda_2$ = 2240 nm while maintaining a high absorptance peak of A ($\lambda_2$ = 2240 nm) = 0.97. On this state, the GST layer is still amorphous and the $VO_2$ NPs turn into tetragonal state. Then continuing to increase the temperature to 180 °C causes the resonant peak to spectrally redshifted to $\lambda_3$ = 2890 nm with the A ($\lambda_3$ = 2890 nm) = 0.99 due to the crystallization of $Ge_2Sb_2Te_5$ layer. Now, decreasing the temperature to 30 °C, causes the resonant peak to blueshift to $\lambda_4$ = 2670 nm with the A ($\lambda_4$ = 2670 nm) = 0.97. The simulated results show nearly perfect absorption for all four possible states of the device. Moreover, the resonant wavelengths can be tuned across a broad 1000 nm spectral band.

The measured absorption spectrum of the device is presented in Fig. 2b. The A ($\lambda$)> 0.9 for all four possible states of the device, and as per the simulated spectra, the absorption peaks can be programmed across a broad 1000 nm band using heat stimuli. The slight difference between the simulation and measurements is likely due to surface roughness and fabrication imperfections, which were included in the simulation. Nonetheless, the main features of the simulation agree with the measurement, and this validates trustworthiness of our design approach.

From the measured spectrum, we can also see that the device can be used as an IR multi-level switch for a fixed wavelength of light. As examples Figs. 2c and 2d show A($\lambda$=1900) and A ($\lambda$=3300 nm) as a function of temperature. With temperature changes, A($\lambda$) changes 80% and 60% at the wavelength of 1900 nm and 3300 nm, respectively. The large absorption contrast is beneficial for the active tunable photonics devices.

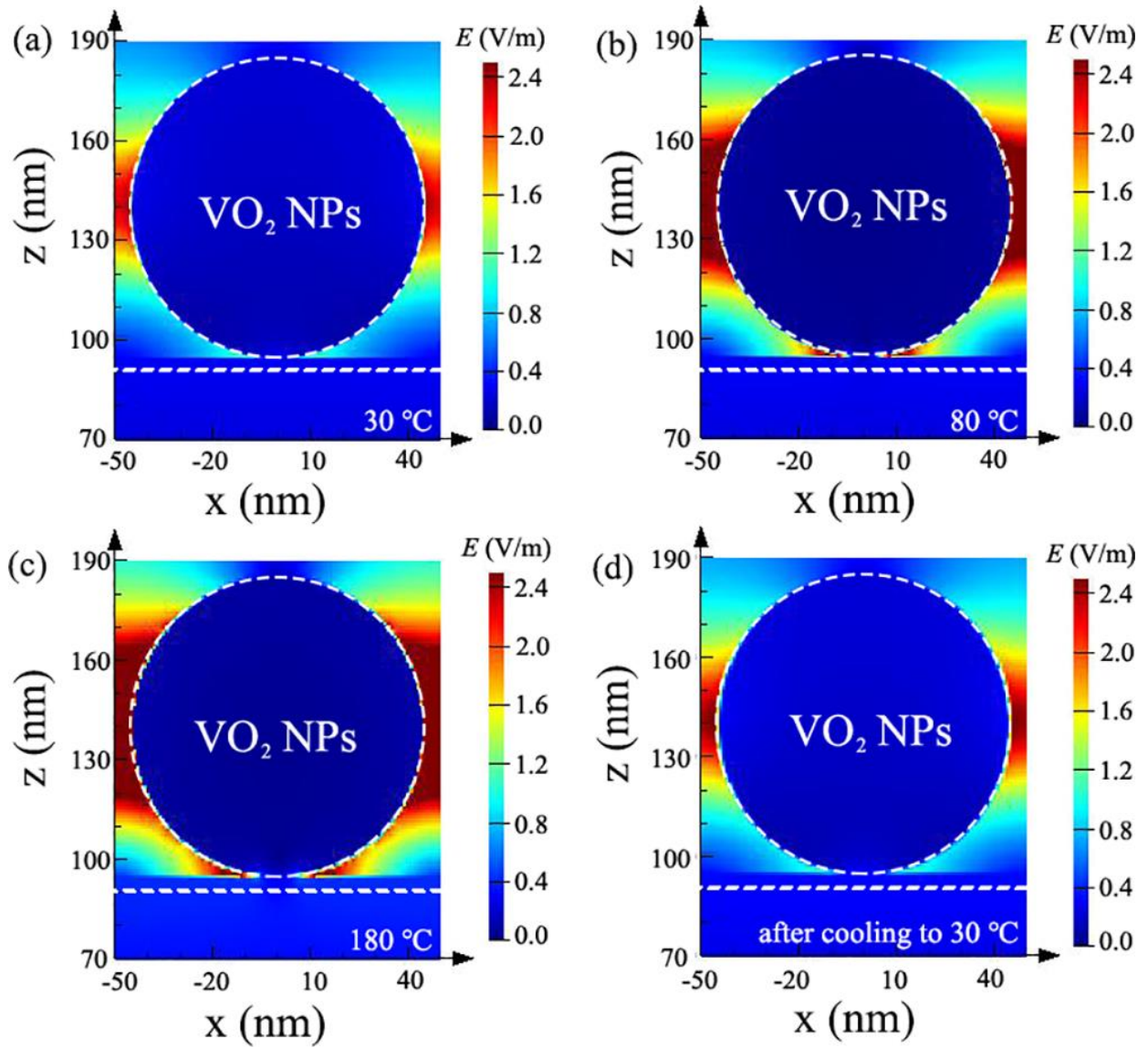

Figure 3 E-field intensity at the resonant wavelength along the cross-section plane for the phase-change metasurfaces at the absorption peak at different temperature: (a) 30 °C; (b) 60 °C; (c) 180 °C; (d) after cooling back to 30 °C.

We have demonstrated 4-wavelength spectral tunability and nearly perfect absorption using our $Ge_2Sb_2Te_5$-$VO_2$ metasurface. We have also simulated the spectral response using the FTDT method to numerically solve Maxwell's equations. Since the simulated spectrum essentially agrees with the measured spectrum, we can trust that FDTD models can be used to explore the underlying physics of the structure. The $VO_2$ NPs (tetragonal state) produce electric dipolar and the strong magnetic resonances between the Au layer and $VO_2$ NPs. These resonances depend on the state and concomitant permittivity of the

VO$_2$ NPs and the Ge$_2$Sb$_2$Te$_5$ layer. To understand how the LSPRs and the phase transition of these materials effect the resonant modes, total electrical (E-) field distribution (E=$\sqrt{|E_x|^2+|E_y|^2+|E_z|^2}$) for the different resonant wavelength was studied. The total E-fied distribution is presented in Fig. 3. At room temperature and at $\lambda_1$= 1841 nm Fig. 3a shows that the E-field between the monoclinic VO$_2$ NPs is weak. Fig. 3b shows the E-field distribution at $\lambda_2$= 2240 nm when the temperature is above 80 °C. At this state, strong electrical resonances were produced in the gap between the tetragonal VO$_2$ NPs and the substrate. However, compared with crystalline Ge$_2$Sb$_2$Te$_5$ layer (Fig. 3c), the amorphous Ge$_2$Sb$_2$Te$_5$ layer decreases the plasmonic resonance between the VO$_2$ NPs and the Au layer. As shown in Fig. 3c, a strong E-field appears not only in the gap of the VO$_2$ NPs, but also at the bottom side of NP when $\lambda_3$= 2890 nm and the temperature increase above 180 °C. At this temperature, the VO$_2$ NPs are in tetragonal state and the GST layer is in its higher refractive index FCC state. Plasmonic resonances are now easily produced between the VO$_2$ NPs and the Au layer. When the temperature cools to room temperature, the E-field distribution the absorption peak shifts to a wavelength of $\lambda_4$= 2670 nm, as seen in Fig. 3d. At this temperature, the VO$_2$ NPs switch into the insulator state yet the Ge$_2$Sb$_2$Te$_5$ remains in the FCC state. Now, it is hard to see any plasmonic effects for the monoclinic VO$_2$ NPs. These field simulation shows how the states changes the Ge$_2$Sb$_2$Te$_5$ layer and the VO$_2$ NPs produce different distributions plasmonic fields, this enabling tuning of the resonant wavelengths and modes.

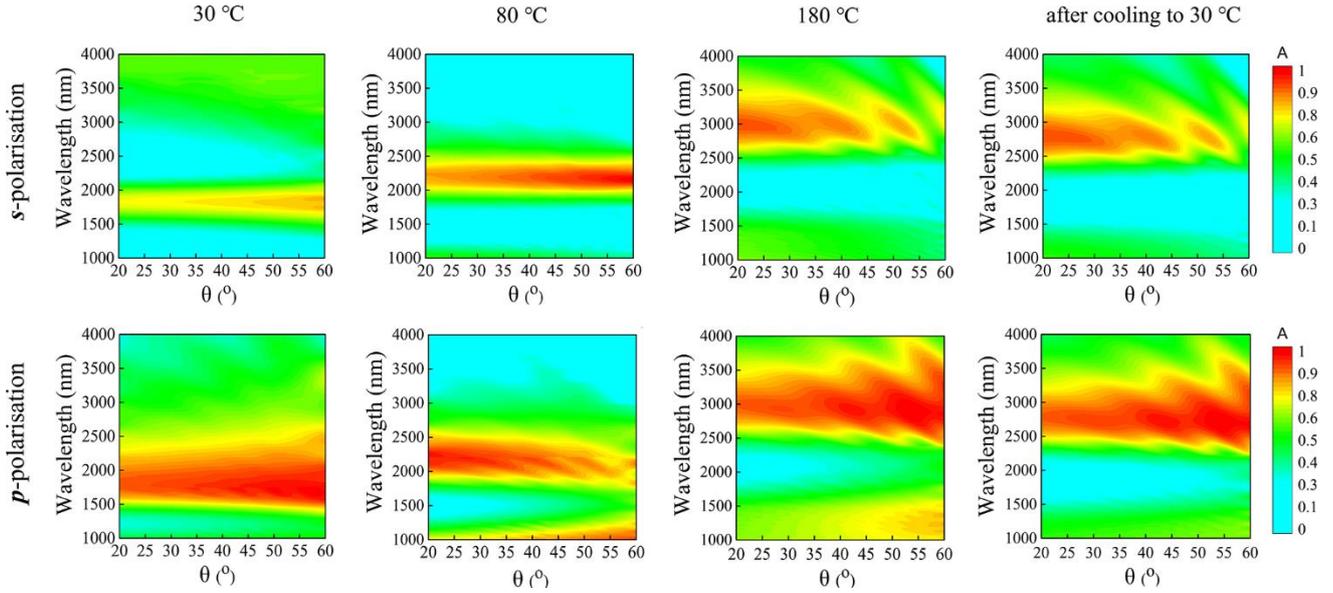

Figure 4 Angular-dependent absorptance spectra at different temperature under the illumination of (a) p-polarization; (b)s- polarization

The proposed metasurface is omnidirectional and polarization independent across a broad range of incident angles. This is a key factor for ideal emitters, photo detectors, and thermal photovoltaics. Fig. 4 shows the A(λ) spectra as the function of incident angle (θ) for p- and s- polarization states. These absorption measurements were measured for the four states of the device. For all possible structural states of the phase change materials and for all the measured angles of incidence, we can see that the absorption is greater than 80% and 90% for the s- and p-polarization states respectively. Between the p- and s-polarized light, the A(ω) peaks exhibit a good spectral overlap for all the four levels.

In summary, we have designed, simulated, and characterized a prototype 4-level reporgrammable near-perfect absorber that exploits the tunable refractive indices of $VO_2$ NPs and $Ge_2Sb_2Te_5$ thins films. The absorber was fabricated using an efficient low cost and lithography-free method. We theoretically and experimentally demonstrated that the peak absorption wavelengths of this perfect absorber could be tuned across a broad wavelength band ranging from 1900 to 2900 nm with the A(ω)>90%. The absorption seems to be insensitive to the polarization state and the angle of incidence. These attributes are attractive for tunable-wavelength omnidirectional emitters and detectors.

**Acknowledgements**

This work was supported by National Research Foundation, Prime Minister's Office, Singapore under its Campus for Research Excellence and Technological Enterprise (CREATE) program and Singapore Ministry of Education (MOE) Academic Research Fund Tier 1 RG103/19 and RG86/20. RES acknowledges support the support granted for the Nanospatial Light Modulators for Next-Gen Display Technologies (NSLM) project (A-Star AME grant number: A18A7b0058).

**Data Availability**

The data that support the findings of this study are available from the corresponding author upon reasonable request.